\RequirePackage{fix-cm}
\documentclass[smallextended]{svjour3}       
\smartqed  
\usepackage{graphicx}
\journalname{Journal of Low Temperature Physics}

\begin{document}

\title{Magnetization process and magnetocaloric effect of the spin-1/2 XXZ Heisenberg cuboctahedron\thanks{This work was financially supported by the grant of The Ministry of Education, Science, Research and Sport of the Slovak Republic under the contract No. VEGA 1/0331/15, by grant of the Slovak Research and Development Agency provided under contract No. APVV-14-0073, as well as, by the internal grant of Faculty of Science of P. J. \v{S}af\'arik University under contract No. VVGS-PF-2016-72606.}}
\titlerunning{Magnetization and magnetocaloric data of the XXZ Heisenberg cuboctahedron}   
\author{Katar\'ina Kar\v{l}ov\'a \and Jozef Stre\v{c}ka}
\institute{Katar\'ina Kar\v{l}ov\'a \at
           Institute of Physics, Faculty of Science of P. J. \v{S}af\'arik University, \\ 
           Park Angelinum 9, 040 01 Ko\v{s}ice, Slovak Republic \\                         
           \email{katarina.karlova@student.upjs.sk}           
           \and
           Jozef Stre\v{c}ka \at
           Institute of Physics, Faculty of Science of P. J. \v{S}af\'arik University, \\ 
           Park Angelinum 9, 040 01 Ko\v{s}ice, Slovak Republic \\
           \email{jozef.strecka@upjs.sk}}

\date{Received: date / Accepted: date}

\maketitle

\begin{abstract}
Magnetic properties of the spin-1/2 XXZ Heisenberg cuboctahedron are examined using exact numerical diagonalization depending on a relative strength of the exchange anisotropy. While the Ising cuboctahedron exhibits in a low-temperature magnetization curve only one-third magnetization plateau, the XXZ Heisenberg cuboctahedron displays another four intermediate plateaux at zero, one-sixth, one-half and two-thirds of the saturation magnetization. The novel magnetization plateaux generally extend over a wider range of magnetic fields with increasing of a quantum ($xy$) part of the XXZ exchange interaction. It is shown that the XXZ Heisenberg cuboctahedron exhibits in a vicinity of all magnetization jumps anomalous thermodynamic behavior accompanied with an enhanced magnetocaloric effect.

\end{abstract}

\section{Introduction}
\label{intro} 

Exotic low-temperature properties of geometrically frustrated quantum spin systems in zero, one, two and three dimensions have attracted appreciable interest in recent years \cite{lacr11,furr13}. An experimental realization of the antiferromagnetic spin-1/2 Heisenberg model on kagom\'e lattice was long sought after as a prominent example of the geometrically frustrated spin system with a spin-liquid ground state \cite{gree01,misg08,zhou16}. This fact has also stimulated particular attention to the antiferromagnetic spin-1/2 Heisenberg cuboctahedron and icosidodecahedron, which can be alternatively viewed as zero-dimensional analogs of the kagom\'e lattice packed on a sphere \cite{rous08,schm05,schn01,schn10}. In addition, the spin-1/2 Heisenberg cuboctahedron as one of the archimedean solids has recently found an intriguing experimental realization through the high-nuclear spin cluster with a Cu$_{12}$La$_8$ core \cite{blak97}. From the theoretical point of view, the magnetization process of the spin-1/2, spin-1, spin-3/2 and spin-5/2 Heisenberg cuboctahedron was investigated in Refs. \cite{rous08,schm05}, while the magnetocaloric effect of the spin-1/2 and spin-1 Heisenberg cuboctahedron was examined in Ref. \cite{schn07,hone08}. It has been evidenced that a height of the last step to the saturation magnetization is twice as large as the other magnetization steps, which can be explained by the highest possible number of localized magnons trapped on square faces \cite{schn01,schu02}. Magnetic properties of the irregular spin-1/2 Heisenberg cuboctahedron with a flat random distribution were studied in Ref. \cite{schn09}. 

It is worthwhile to remark that the antisymmetric Dzyaloshinsky-Moriya term \cite{moriya} as a possible source of the magnetic anisotropy can be generally ruled out due to an anticipated presence of the inversion center in the middle of each bond of the cuboctahedron. In this regard, the XXZ exchange anisotropy  represents for the Heisenberg cuboctahedron the leading-order term determining the magnetic anisotropy on assumption that a local crystal field of negatively charged ligands around each magnetic center has a unique anisotropy axis \cite{jough}. In the present work, we will therefore examine an influence of the exchange anisotropy on a low-temperature thermodynamics of the spin-1/2 XXZ Heisenberg cuboctahedron, which has not been dealt with yet. 

\section{Model and methods}
\label{model}

Let us consider spin-1/2 XXZ Heisenberg cuboctahedron (see Fig. \ref{fig1}) given by the Hamiltonian
\begin{eqnarray}
\hat{\cal H} = J \sum_{\langle i,j \rangle}^{N_b} \left[\Delta \left(\hat{S}_{i}^x\hat{S}_{j}^x + \hat{S}_{i}^y\hat{S}_{j}^y \right) + \hat{S}_{i}^z\hat{S}_{j}^z\right] - h \sum_{i=1}^N \hat{S}_i^z,
\label{ham}
\end{eqnarray}
where $\hat{ S}_i^\alpha$ denotes spatial projections ($\alpha=x, y, z$) of the spin- 1/2 operator placed at $i$th  vertex of the cuboctahedron, the first summation accounts for the antiferromagnetic interaction $J>0$ between the nearest-neighbor spins and $\Delta \in \langle 0; 1\rangle$ is the anisotropy parameter in the XXZ Heisenberg interaction. Two limiting cases $\Delta=0$ and $\Delta=1$  correspond to the Ising and isotropic Heisenberg models, respectively. The second summation accounts for the Zeeman's energy of magnetic moments in the external magnetic field $h > 0$ and finally, $N$ ($N_b$) denotes the total number of spins (bonds) of the cuboctahedron. To obtain exact diagonalization data for the magnetization and entropy of the spin-1/2 XXZ Heisenberg cuboctahedron we have adapted the subroutine edfulldiag from Algorithms and Libraries for Physics Simulations (ALPS) project \cite{bau11}. This approach allows a rigorous calculation of the magnetization process and magnetocaloric properties of the model under investigation, which will be the central issue of our subsequent analysis. 

\begin{center}
\hspace{2cm}
\begin{figure}
\includegraphics[width=0.6\textwidth]{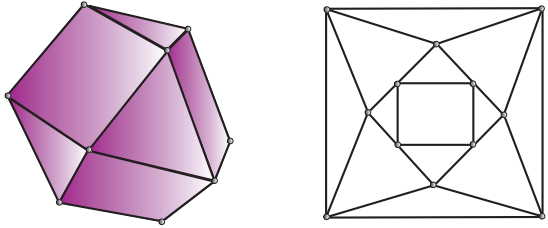}
\caption{The cuboctahedron and its planar projection. (Color figure online)}
\label{fig1}       
\end{figure}
\end{center}

\section{Results and discussion}
\label{result}

\begin{figure}
\includegraphics[width=0.7\textwidth]{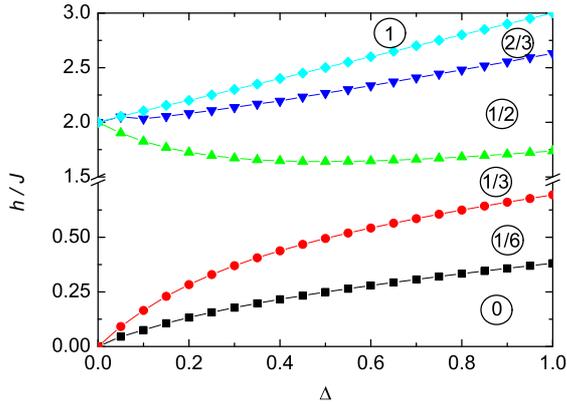}
\vspace*{-0.6cm}
\caption{The ground-state phase diagram of the spin-1/2 XXZ Heisenberg cuboctahedron 
in the $\Delta - h/J$ plane. Acronyms in round circles determine the total magnetization 
normalized with respect to its saturation value. (Color figure online)}
\label{fig2}       
\end{figure}

Let us start our discussion with the ground-state diagram of the spin-1/2 XXZ Heisenberg cuboctahedron, which is depicted in Fig. \ref{fig2} in the $\Delta - h/J$ plane. Each area delimited by displayed phase boundaries corresponds to a different ground state, which is distinguished according to the corresponding value of the total magnetization normalized with respect to its saturation value. With exception of a trivial fully polarized ground state, the spin-1/2 XXZ Heisenberg cuboctahedron exhibits for any $\Delta \neq 0$ another five ground states, which should be manifested in a zero-temperature magnetization curve as intermediate plateaux at zero, one-sixth, one-third, one-half and two-thirds of the saturation magnetization. This result is in sharp contrast with much simpler behavior of the Ising cuboctahedron retrieved in the limiting case $\Delta = 0$, which exhibits in a zero-temperature magnetization process just a single one-third plateau before the magnetization reaches saturation. It is noteworthy that only the state corresponding to a five-sixths intermediate plateau is not realized as a particular ground state out of all available lowest-energy states of the spin-1/2 XXZ Heisenberg cuboctahedron with the total spin $S_T=0,1,\ldots 6$, which can be attributed to the highest possible number of localized magnons trapped on two opposite square faces of the cuboctahedron (see Fig. \ref{fig1}) within the two-thirds magnetization plateau \cite{schn01}.

\begin{figure}
\hspace{-0.7cm}
\includegraphics[width=0.6\textwidth]{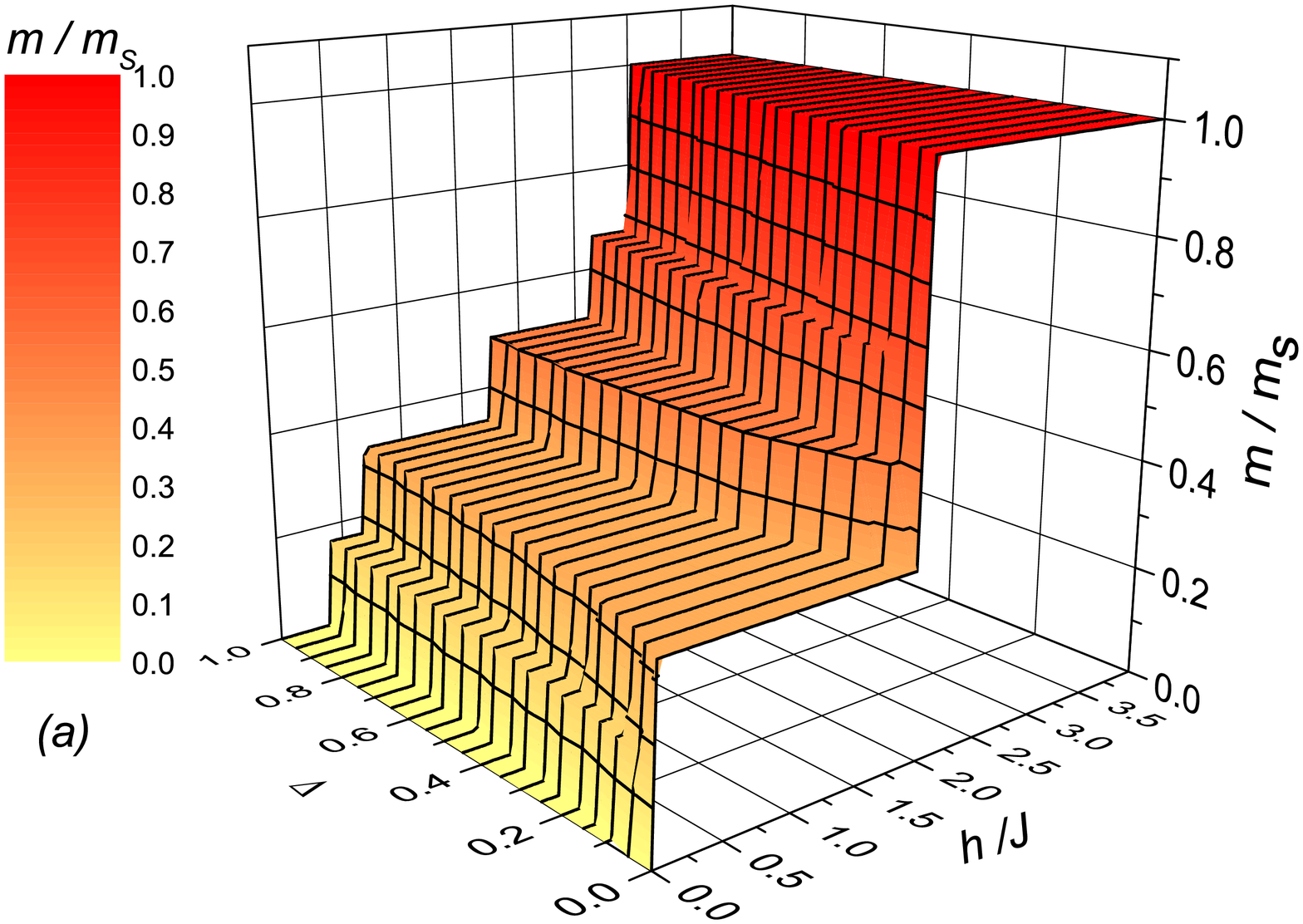}
\hspace{-0.7cm}
\includegraphics[width=0.6\textwidth]{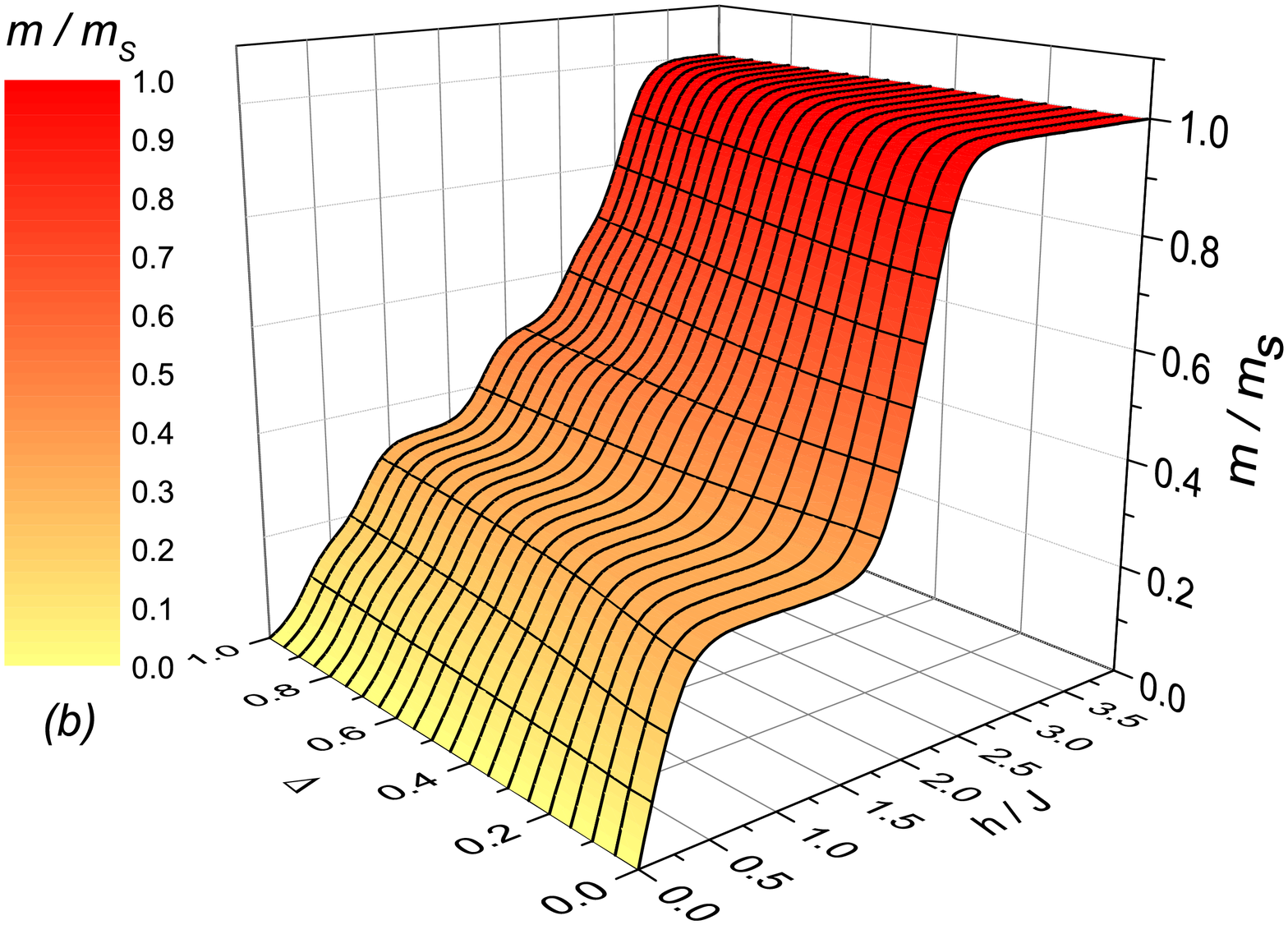}
\vspace*{-0.8cm}
\caption{3D plot of the magnetization of the spin-1/2 XXZ Heisenberg cuboctahedron as a function of the external magnetic field $h/J$ and exchange anisotropy $\Delta$ at two different temperatures: (a) $k_{\rm B} T/J = 0.001$; (b) $k_{\rm B} T/J = 0.1$. (Color figure online)}
\label{fig3}       
\end{figure}

To bring an insight into how the particular ground states are manifested at finite temperatures we depict in Fig. \ref{fig3} three-dimensional (3D) plot of the magnetization as a function of the magnetic field and anisotropy parameter $\Delta$ at the sufficiently low ($k_{\rm B} T/J = 0.001$) and moderate ($k_{\rm B} T/J = 0.1$) temperatures. Although there are no true magnetization plateaux and jumps at any finite temperature, the low-temperature magnetization curve displayed in Fig. \ref{fig3}(a) is strongly reminiscent of the actual magnetization plateaux and jumps observed strictly at zero temperature. Moreover, it can be actually seen from Fig. \ref{fig3}(a) that only one intermediate plateau at one-third of the saturation magnetization is realized in the Ising limit ($\Delta =0$), while the spin-1/2 XXZ Heisenberg cuboctahedron generally exhibits for any $\Delta \neq 0$ another four intermediate plateaux at zero, one-sixth, one-half and two-thirds of the saturation magnetization. The novel magnetization plateaux gradually extend over a wider range of the magnetic fields with increasing of a quantum ($xy$) part of the XXZ exchange interaction at the expense of the original one-third and saturation plateaux. It can be found that the one-third plateau is approximately $35\%$ narrower for the isotropic Heisenberg cuboctahedron ($\Delta=1$) in comparison with the Ising cuboctahedron ($\Delta=0$). Finally, it is worthwhile to remark that the five-sixths plateau is indeed absent in the magnetization curve of the spin-1/2 XXZ Heisenberg cuboctahedron regardless of the exchange anisotropy \cite{schn01}.

For a comparison, the magnetization process of the spin-1/2 XXZ Heisenberg cuboctahedron at the moderate temperature $k_{\rm B} T/J = 0.1$ is shown in Fig. \ref{fig3}(b). As one can see, the magnetization plateaux and jumps become gradually smoother upon rising temperature. Actually, it can be found from Fig. \ref{fig3}(b) that all magnetization plateaux except one-third plateau  completely disappear from the relevant magnetization curve provided that the anisotropy parameter $\Delta$ is sufficiently small (i.e. near the Ising limit $\Delta=0$). On the other hand, two widest one-third and one-half magnetization plateaux can be still discerned at this temperature in the magnetization curve close to the isotropic Heisenberg limit ($\Delta=1$), while other narrower magnetization plateaux are reflected merely as less pronounced inflection points.

\begin{figure}
\hspace{-0.7cm}
\includegraphics[width=0.6\textwidth]{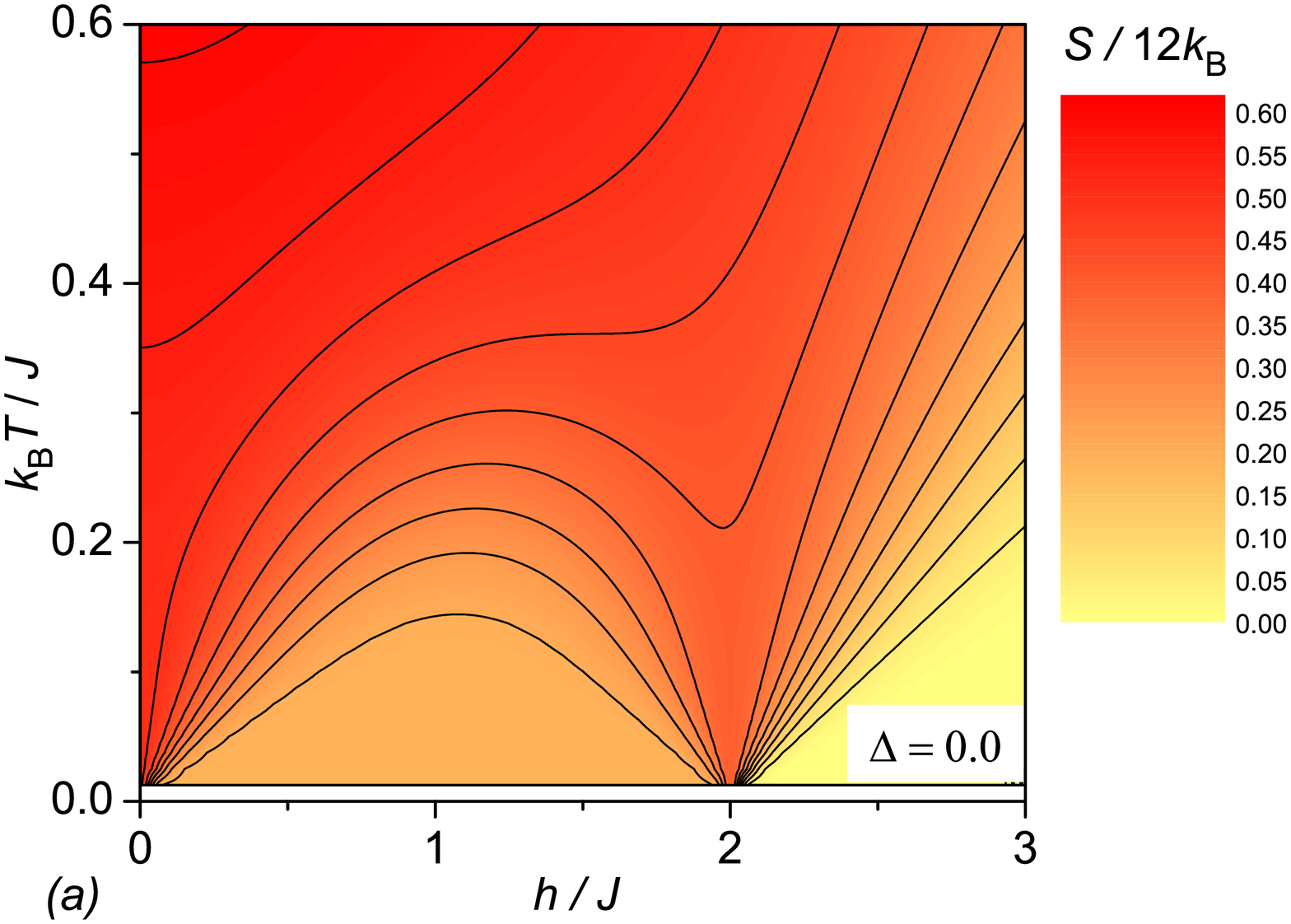}
\hspace{-1.2cm}
\includegraphics[width=0.6\textwidth]{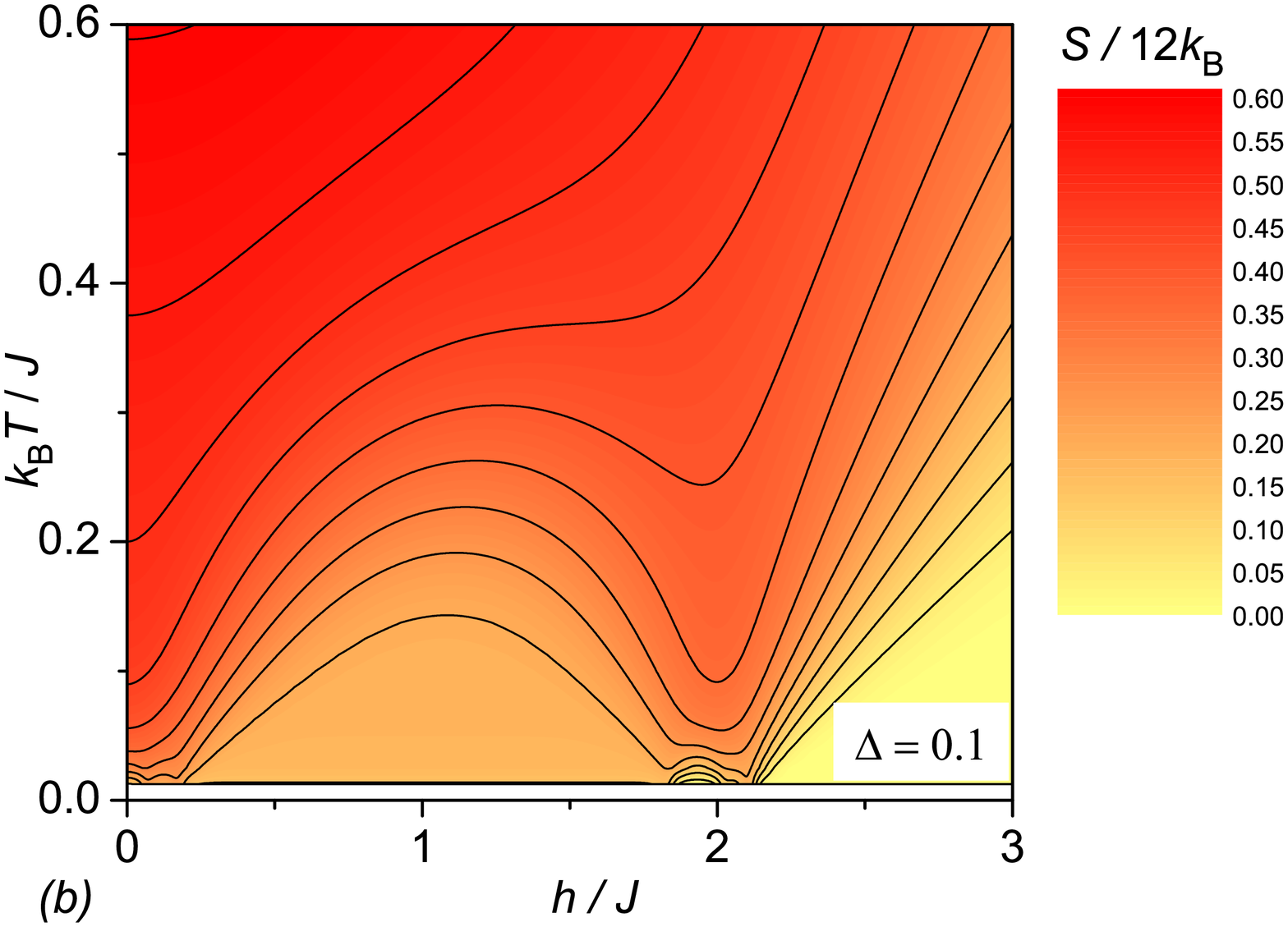}
\vspace*{-0.7cm}
\caption{A density plot of the entropy per one spin of the spin-1/2 XXZ Heisenberg cuboctahedron as a function of the magnetic field and temperature for two different values
of the anisotropy parameter: (a) $\Delta=0.0$; (b) $\Delta=0.1$. (Color figure online)}
\label{fig4}       
\end{figure}

Next, let us also examine magnetocaloric properties of the spin-1/2 XXZ Heisenberg cuboctahedron, which can be particulary interesting especially in a vicinity of the magnetization jumps. First, we will discuss the Ising limiting case with $\Delta=0$, for which a few isentropy lines are plotted in Fig. \ref{fig4}(a) in the magnetic field-temperature plane. The displayed isentropy lines can also be viewed as a temperature response with respect to varying external magnetic field during the adiabatic demagnetization. The Ising cuboctahedron exhibits a giant magnetocaloric effect just above (below) of two magnetization jumps, where a sharp increase (decrease) of temperature is invoked upon lowering of the magnetic field. An absence of zero magnetization plateau in a magnetization process of the Ising cuboctahedron causes a giant magnetocaloric effect in a vicinity of zero field, which makes this frustrated spin structure quite promising refrigerant enabling cooling down to absolute zero temperature quite similarly to the Ising octahedron and dodecahedron  \cite{stre15}. 

\begin{figure}
\hspace{-0.7cm}
\includegraphics[width=0.6\textwidth]{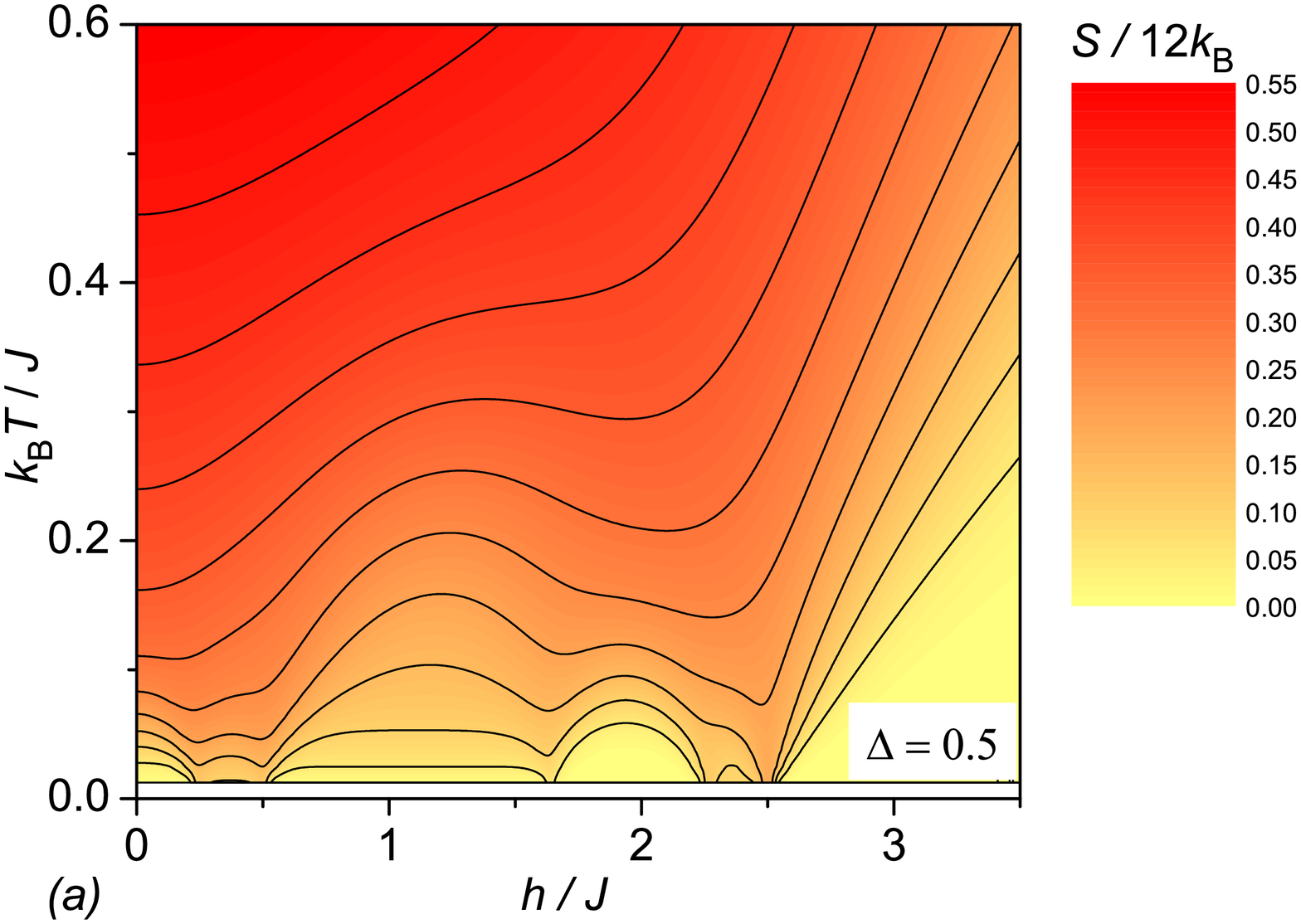}
\hspace{-1.2cm}
\includegraphics[width=0.6\textwidth]{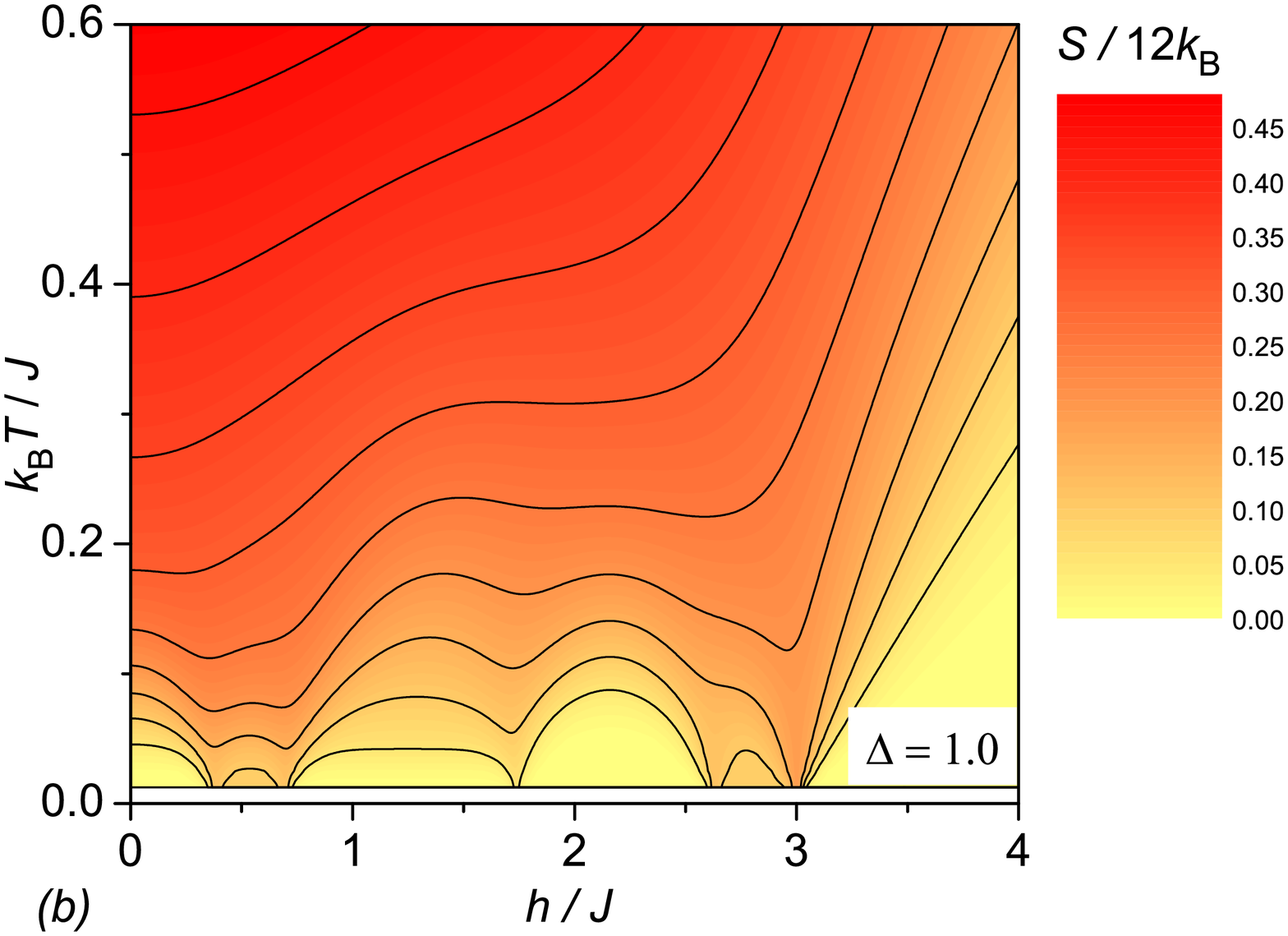}
\vspace*{-0.7cm}
\caption{A density plot of the entropy per one spin of the spin-1/2 XXZ Heisenberg cuboctahedron as a function of the magnetic field and temperature for two different values of the anisotropy parameter: (a) $\Delta=0.5$; (b) $\Delta=1.0$. (Color figure online)}
\label{fig5}       
\end{figure}

The density plot of the entropy is displayed in Fig. \ref{fig4}(b) for relatively low value of the anisotropy parameter $\Delta=0.1$ close enough to the Ising limiting case. Though it is still possible to observe during the adiabatic demagnetization a steep increase (decrease) of temperature just below (above) of the magnetization jumps, the most essential difference is that the temperature shows a less marked decrease (just down to some finite value) near zero field due to a presence of tiny plateau with a zero magnetization. The other intermediate magnetization plateaux occur in a relatively narrow range of magnetic fields for low values of the anisotropy parameter $\Delta$ and hence, the deviation from an anomalous magnetocaloric behavior manifest itself just at relatively low temperatures. 

Contrary to this, the isentropy lines of the isotropic Heisenberg cuboctahedron $\Delta=1$ indicate a remarkable change of temperature in the vicinity of all magnetization steps [see Fig. \ref{fig5}(b)]. It is quite curious that the similar magnetocaloric behavior can be already detected in the spin-1/2 XXZ Heisenberg cuboctahedron with a relatively strong exchange anisotropy $\Delta=0.5$ as evidenced in Fig. \ref{fig5}(a). As a matter of fact, the temperature exhibits in both aforementioned cases a relatively steep decrease (increase) just above (below) each critical field associated with the magnetization jump. From the perspective of the cooling efficiency, the spin-1/2 XXZ Heisenberg cuboctahedron with the exchange anisotropy $\Delta > 0.1$ should be however regarded as much worse refrigerant system, because temperature finally shows a gradual increase during the adiabatic demagnetization in a close vicinity of zero magnetic field due to a presence of zero magnetization plateau.  

\section{Conclusion}
\label{conc}
The present work deals with the magnetization process and magnetocaloric properties of the spin-1/2 XXZ Heisenberg cuboctahedron, which have been analyzed by means of the exact diagonalization method. It has been demonstrated that the magnetization curve of the spin-1/2 XXZ Heisenberg cuboctahedron posseses several intermediate plateaux and an anomalous magnetocaloric effect close to respective magnetization steps. While the spin-1/2 Ising cuboctahedron exhibits only one intermediate plateau at one-third of its saturation value, the spin-1/2 XXZ Heisenberg cuboctahedron additionally exhibits other four intermediate plateaux of a quantum nature. An enhanced  magnetocaloric effect has been detected in a vicinity of all magnetization steps. The magnetocaloric effect of the spin-1/2 XXZ Heisenberg cuboctahedron at the anisotropy parameter $\Delta=0.1$ behaves quite similarly as the Ising limiting case $\Delta=0$. On the other hand, the magnetocaloric effect of the spin-1/2 XXZ Heisenberg cuboctahedron for the anisotropy parameter $\Delta > 0.5$ behaves quite similarly as the isotropic Heisenberg case $\Delta=1$.


\begin{thebibliography}{20}
\bibitem{lacr11} C. Lacroix, Ph. Mendels, F. Mila, Introduction to Frustrated Magnetism. Springer, Heidelberg (2011).
\bibitem{furr13} A. Furrer, O. Waldmann, \textit{Rev. Mod. Phys.}, \textbf{85}, 367-420 (2013).
\bibitem{gree01} J. E. Greedan, \textit{J. Mater. Chem.}, \textbf{11}, 37-53 (2001).
\bibitem{misg08} G. Misguich, Quantum Spin Liquids, in Exact Methods in Low-Dimensional Statistical Physics and Quantum Computing, eds. J. Jacobsen, S. Ouvry, V. Pasquier, D. Serban, L. F. Cugliandolo. Oxford University Press, Oxford (2008).
\bibitem{zhou16} Y. Zhou, K. Kanoda, T.-K. Ng, preprint arxiv: 1607.03228
\bibitem{rous08} I. Rousochatzakis, A. M. L\"auchli, F. Mila, \textit{Phys. Rev. B}, \textbf{77}, 094420 (2008).
\bibitem{schm05} R. Schmidt, J. Richter, J. Schnack, \textit{J. Magn. Magn. Mater}, \textbf{295}, 164-167 (2008).
\bibitem{schn01} J. Schnack, H.-J. Schmidt, J. Richter, J. Schulenburg,  \textit{Eur. Phys. J. B}, \textbf{24}, 475-481 (2001).
\bibitem{schn10} J. Schnack, \textit{Dalton Trans.}, \textbf{39}, 4677 (2010).
\bibitem{blak97} A. J. Blake, R. O. Gould, C. M. Grant, P. E. Y. Milne, S. Parsons, R. E. P. Winpenny, \textit{J. Chem. Soc., Dalton Trans.}, 485-495 (1997).
\bibitem {schn07} J. Schnack, R. Schmidt, J. Richter, \textit{Phys. Rev. B}, \textbf{76}, 054413 (2007).
\bibitem{hone08} A Honecker, M. E. Zhitomirsky \textit{J. Phys.: Conf. Ser.}, \textbf{145}, 012082 (2009).
\bibitem{schu02} J. Schulenburg et al., \textit{Phys. Rev. Lett.}, \textbf{88}, 167207 (2002).
\bibitem{schn09} J. Schnack, R. Schnalle, \textit{Polyhedron}, \textbf{28}, 1620-1623 (2009).
\bibitem{moriya} T. Moriya, \textit{Phys. Rev.}, \textbf{120}, 91 (1960).
\bibitem{jough} L. J. de Jough, A. R. Miedema \textit{Adv. Phys.}, \textbf{23}, 1 (1974).
\bibitem{bau11} B. Bauer, L.D. Carr, H.G. Evertz, A. Feiguin, J. Freire, S. Fuchs, L. Gamper, J. Gukelberger, E. Gull, S. Guertler, A. Hehn, 
R. Igarashi, S.V. Isakov, D. Koop, P.N. Ma, P. Mates, H. Matsuo, O. Parcollet, G. Pawlowski, J.D. Picon, L. Pollet, E. Santos, V.W. Scarola, 
U. Schollw\"ock, C. Silva, B. Surer, S. Todo, S. Trebst, M. Troyer, M.L. Wall, P. Werner, S. Wessel, \textit{J. Stat. Mech.: Theor. Exp.}, \textbf{2011}, P05001 (2011).  
\bibitem{stre15} J. Stre\v{c}ka, K. Kar\v{l}ov\'a, T. Madaras, \textit{Physica B}, \textbf{466}, 76-85 (2015).
\end{thebibliography}
\end{document}